\documentclass[12pt,preprint]{aastex}

\slugcomment {}

\shorttitle{TW Hya Association Membership and New WISE-detected Circumstellar Disks}
\shortauthors{Schneider et al.}

\begin{document}

\title{TW Hya Association Membership and New WISE-detected Circumstellar Disks}

\author{Adam Schneider}
\affil{Department of Physics and Astronomy, University of Georgia,
    Athens, GA 30602}
\email{aschneid@physast.uga.edu}

\author{Carl Melis}
\affil{Center for Astrophysics and Space Sciences, University of California,
    San Diego, CA 92093}
\email{cmelis@ucsd.edu}

\and

\author{Inseok Song}
\affil{Department of Physics and Astronomy, University of Georgia,
    Athens, GA 30602}
\email{song@physast.uga.edu}

\begin{abstract}
We assess the current membership of the nearby, young TW Hydrae Association and examine newly 
proposed members with the Wide-field Infrared Survey Explorer (WISE) to search for infrared excess 
indicative of circumstellar disks.  Newly proposed members TWA 30A, TWA 30B, TWA 31, and TWA 32 all show 
excess emission at 12 and 22 $\mu$m providing clear evidence for substantial dusty circumstellar  
disks around these low-mass, $\sim$8 Myr old stars that were previously shown to likely be accreting 
from circumstellar material.  TWA 30B shows large amounts of self-extinction, likely due to an edge-on 
disk geometry.  We also confirm previously reported circumstellar disks with WISE, and determine a 
22 $\mu$m excess fraction of $42^{+10}_{-9}$\% based on our results.           
\end{abstract}


\section{Introduction}
Young stars in the vicinity of the classical T-Tauri star TW Hydrae were first proposed to be associated 
by \cite{kas97}.  Since then, many new members have been proposed and studied, and the TW 
Hyadrae association (TWA) is now currently made up of a few dozen members (see \citealt{bar06} 
and \citealt{tor08} for the most recent membership reviews).  TWA has an age of $\sim$8 Myr, a critical 
age of stellar evolution and planet formation, between that of T-Tauri and main sequence stars.  This 
association's young age and proximity to the Sun ($\sim$50 pc) have made it an area of intense study, 
most notably for low-mass members (such as 2M1207b; \citealt{cha04}) and circumstellar disks (e.g. 
\citealt{wei04}, \citealt{low05}, \citealt{reb08}, and \citealt{riaz08}).  

Previous studies of dusty disks in TWA have shown a diverse array of disk properties for its members, 
spanning a range of evolutionary stages.  The first four disks (TW Hya, Hen-3 600, HD 98800, and 
HR4796A) were detected with the {\it Infrared Astronomical Satellite} (IRAS).  After the discovery of 
new members, surveys were performed to identify any additional sources with IR-excess above 
predicted stellar photospheres using mid-IR imaging from the ground \citep{wei04} and space 
\citep{low05} with the {\it Spitzer Space Telescope} \citep{wer04}.  Until recently, no additional 
members showing strong IR-excess emission have been found, though some members showing small 
amounts of excess ($\tau$ $\equiv$ $L_{IR}$/$L_*$ $<$ 0.5\%) have been proposed (TWA 7 and TWA 
13 - \citealt{low05}; TWA 8B and TWA 19 - \citealt{reb08}).  The surveys by \cite{wei04} and \cite{low05} 
show a clear bimodal distribution of excess emission at mid-IR wavelengths, indicating that most 
members have negligible amounts of warm dust surrounding them.  Disks have since been discovered 
around 4 new TWA members not included in the previous mid-IR studies, 2M1207 \citep{ste04}, SSPM 
1102 (\citealt{riaz08} and \citealt{mor08}), TWA 30A \citep{loop10a}, and TWA 30B \citep{loop10b}.  
These four most recent disk discoveries are all around late-type members (spectral types $\geq$ M4), 
indicating a very high disk fraction for objects of this type in TWA.

A new method of identifying young stars employed by \cite{rod11} utilizing UV-excess measurements
from the {\it Galaxy Evolution Explorer} (GALEX; \citealt{mar05}) led to the proposal of several new TWA 
candidates. \cite{shk11}, using GALEX in a similar way, independently confirm the membership of one 
of the proposed candidates of \cite{rod11} (TWA 32) and propose another additional member (TWA 31).

With the {\it Wide-Field Infrared Explorer} (WISE), we now have the opportunity to explore each current 
member, old and new, with mid-IR wavelengths to search for excess emission indicative of circumstellar 
material.  In this letter, we summarize assessment of TWA membership, and attempt to re-analyze all 
previous reports of IR-excess. We also evaluate all members for which mid-IR photometry has not been 
performed, including new members, to search for evidence of any surrounding dusty disk.

\section{Current Membership of TWA}
Establishing the membership status of TWA is difficult.  This is partly due to the overlap with the 
Lower-Centaurus Crux (LCC) region of the Scorpius Centaurus complex (Sco-Cen).  Young stars found 
in this area can be difficult to place because the age of the LCC ($\sim$10-20 Myr) is similar enough to 
that of TWA that youth indicators alone cannot distinguish between the two.  Typically, a distance 
measurement is relied upon to make the final distinction, because the LCC is generally further away 
than TWA (LCC $\sim$120 pc; TWA $\sim$50 pc).  However, with the addition of new members, such as 
TWA 29 (d $\sim$90 pc) and TWA 31 (d $\sim$110 pc), the boundary between the LCC and TWA is 
becoming much more ambiguous, providing evidence that TWA may indeed be the front edge of the 
LCC, as proposed by \cite{song03}.  For the purposes of this work, we choose a distance cut of 100 pc 
for the identification of TWA members, acknowledging that this choice will need further revision if more 
members are identified at intermediate distances.         

A current summary of the properties of proposed TWA members is given in Table 1.  When more than 
one value for a property of a particular object is available, we opt for the most recent measurement.  A 
complete list of members is crucial when evaluating properties such as excess fractions, age, and group 
velocities.  \cite{tor08} evaluated the membership of all proposed TWA members up to TWA designation 
28 (including eight objects labeled as secondaries to TWA members), with the exception of 4 possible 
members for which kinematical data was found to be lacking.  By applying a convergence method to the 
remaining 32 stars, 6 were found to have a TWA membership probability of zero (15A, 17, 18, 19A, 19B, 
and 24), and we agree with this assessment based on the discrepant distances found for these stars (d 
$>$ 100) compared to those of members found to have a high probability of membership (70 
$>$ d $>$ 25).  Of the remaining, 2 stars were categorized as unlikely members (12 and 21), and 2 as 
possible members (6 and 14).  A total of 22 stars were found to be high probability members.  

TWA 15B, TWA 22, TWA 23, and TWA 28 were the four possible members not evaluated by \cite{tor08}.  
As of this writing, TWA 15B still lacks the kinematical information necessary to properly evaluate its 
membership, though its estimated distance (100 pc) from \cite{shk11}, inconsistent with known TWA 
members, leads us to label its status as questionable.  \cite{teix09} performed a thorough analysis of the 
kinematics of TWA 22, but the results were somewhat inconclusive.  We retain TWA 22 as a possible
member.  \cite{shk11} reevaluate the kinematics of TWA 23 (found to be spectroscopic binary), and 
show its UVW space motion is consistent with being a TWA member.  In a similar fashion as TWA 22, 
\cite{teix08} thoroughly evaluate the membership of TWA 28 with high precision proper motion and 
parallax measurements, and determine it is a true member of TWA. 

Since the review of \cite{tor08}, six objects have been assigned TWA designations (11C, 29, 30A, 30B, 
31, and 32).  TWA 11C was discovered and determined to be a member of TWA by \cite{kas08} via its 
common proper motion to TWA 11AB and strong lithium absorption and H$\alpha$ emission confirming 
its youth.  For TWA 29, first proposed by \cite{loop07}, we determine new proper motion values based on 
its 2MASS and WISE positions.  These new values, listed in Table 1, along with its estimated distance, 
strong H$\alpha$ emission, and position relative to other TWA members, make its membership highly
likely.  TWA 30A and TWA 30B have been determined to be highly likely members by \cite{loop10a} and 
\cite{loop10b} respectively, and we agree with their analysis.  \cite{shk11} show that proposed members 
TWA 31 and 32 share  spectroscopic, photometric, and kinematic properties with known members, 
thereby solidifying their membership.  This is definitely the case for TWA 32, with UVW space 
motions\footnote{$UVW$ are defined with respect  to the Sun. U is positive toward the Galactic center, 
V is positive in the direction of Galactic rotation, and W is positive toward the north Galactic pole.} 
consistent with known TWA members.  For TWA 31, though, this is not so obvious.    

\cite{shk11} quote a photometric distance to TWA 31 of 110 pc, a distance inconsistent with that of known 
members.  Additionally, in their analysis, they compare the UVW of TWA 31 to some stars with 
questionable membership, such as TWA 12, 14, and 22, making the UVW for TWA 31 seem more 
reasonable than it should.  The claim that TWA 31 is certainly too young to be an LCC member is valid 
when the LCC is assigned an age of 16 Myr.  However, if the LCC is younger ($\sim$10Myr, as suggested 
by \cite{song12}), then the use of age indicators as a deciding factor for membership in TWA or the LCC 
is unsound.  As mentioned above, this may be one of many possible objects between TWA and the LCC 
that make a distinct boundary between the two unclear.  We retain TWA 31 as a possible member for this 
study because of the large uncertainty ($\sim$10\%) in its distance estimate.     
    
In summary, of all stars with TWA designations, we find there to be 30 bona-fide TWA members, 4 
possible members, one questionable member, 2 unlikely members, and 7 nonmembers.  This is 
displayed in Table 1 with the following symbols representing the various designations: Y = true member,
Y? = possible member, ? = questionable member, N? unlikely member, and N = nonmember.

\section{IR-Excess with WISE}

For every confirmed member of TWA, we cross-correlate with the WISE catalog.  Out of
the 30 confirmed members of TWA, 26 were detected with WISE.  The four not included in WISE 
(TWA 3B, TWA 4B, TWA 5B, and TWA 11B), all companions, are too close to their primary members
to be resolved individually.  Three members were too faint to be detected in WISE channel 4, only 
having upper limits (TWA 9B, TWA 26, and TWA 29).  For these members, we searched the Spitzer
Heritage Archive for possible detections at 24 $\mu$m with the Multiband-Imaging Photometer for 
Spitzer (MIPS; \citealt{rie04}) to contribute to our mid-IR excess statistics.  TWA 26 was the only observed member able to be detected.  
Source extraction was performed on the MIPS Post Basic Calibrated Data (pBCD) mosaic files using
the Astronomical Point Source Extraction (APEX) package for this source, and the resulting flux measurement is
included in Table 2.     

For the remaining members, all were detected 
in all four WISE bands.  Spectral energy distributions (SEDs) were generated using available
catalog data from Tycho-2, 2MASS, and WISE.  Stellar photosphere fluxes are predicted from 
our SED fitting technique (as described in \citealt{rhee07}).  We convert the WISE magnitudes to 
fluxes using the conversion factors from \cite{wri10} for each source, then look for excess by
examining each SED individually and checking the ratio of the WISE channel 4
flux versus the predicted photospheric flux at 22 $\mu$m.  We flag potential excess sources as those
having a WISE measurement greater than 5 times the WISE photometric uncertainty above our photospheric flux estimate at 22 $\mu$m.    
Predicted photospheric flux estimates, measured WISE fluxes, and significance of excess 
calculations for possible and true members of TWA are given in Table 2 for WISE channels 1, 3, 
and 4 (WISE channel 2 is not included because the spectral models in the WISE channel 2 region 
are affected by the fundamental CO bandhead at 4.7 $\mu$m for later-type stars).  The significance of excess 
evaluation only takes into account the WISE photometric uncertainties and not any uncertainty 
from the spectral fitting.  

As seen in Table 2, we recover previous reports of excess indicative of a dusty disk
for TWA 1, 3A, 4A, 7, 11A, 27, 28, 30A, and 30B.  We also find strong excess around
newly proposed members TWA 31 and 32.  The presence of dust around TWA 30A was inferred by \cite{loop10a}, 
and 30B was determined to have IR-excess indicative of a disk in \cite{loop10b} based on their near-IR
spectra, and we show the shape of the mid-IR emission in the SEDs of each in Figure
1, along with the two new disk detections for TWA 31 and 32.  As seen in the figure, the excess mid-infrared 
emission for TWA 30A, 31, and 32 are each fit with a single blackbody. Incomplete excess 
information at longer wavelengths precludes a more detailed model fit. 
From these simple blackbody fits we are able to provide an estimate of 
each disk's characteristic temperature and absorbing area (given by the 
parameters 'Tdust' and '$\tau$', respectively) from the available data. 
These values are displayed in Figure 1.  A color-color diagram showing the clear distinction between 
excess and non-excess stars is shown in Figure 2.   

The SED for TWA 30B shows a drastically different shape than that of the other sources.  We believe that this 
unusual SED is due to the disk inclination.  By using 
a spectral model for an M4 star scaled to the distance measured by \cite{loop10b} for TWA 30B, 
(dashed line in Figure 1), we see that the photospheric emission is obscured at least up to WISE channel 2.  
This object is likely obscured by an edge on disk, as suggested by \cite{loop10b}.  To fit the current data, 
we scale down the model spectrum to meet the J-band flux and fit the excess emission using two
blackbodies of 660 and 190 K located at distances of 2 and 31 AU, respectively.  The inner disk temperature
is in good agreement with that found by \cite{loob10b}.  This is only one potentially 
viable model capable of fitting the currently available data.  A more thorough analysis of this source
across many wavelengths needs to be carried out before a complete understanding can be obtained.

\cite{reb08} reevaluated the work done by \cite{low05} to determine a 24 $\mu$m
excess fraction of 30$\pm$11\% for TWA.  The disk fraction for TWA was again
estimated in \cite{loop10a}, using the list of members from \cite{mam05}. 
They determine an overall disk fraction of $35^{+11}_{-8}$\%.  Using our updated list
of current members (29 total, 23 individually detected at 22 $\mu$m with WISE, and one (TWA 26) detected at
24 $\mu$m with Spitzer), and WISE
photometry, we determine a mid-IR excess fraction of (10/24) $42^{+10}_{-9}$\%.  Members not detected at 22 $\mu$m
with WISE (or 24 $\mu$m with Spitzer) are not included in our excess statistics.  Uncertainties are estimated
following the method found in the appendix of \cite{bur03}.  If possible members
are included, the excess fraction becomes (11/28) $39^{+10}_{-8}$\%.  If
we focus solely on substellar members (spectral type $>$ M6), all three detected in 
WISE channel 4 show excess emission (though TWA 26 shows no excess at 24 $\mu$m).  Comparing this with that of the stellar
members (7/23) $30^{+11}_{-7}$\%, we see a possible indication of a difference in the disk decay lifetimes of stellar
and substellar members of TWA.  Though the sample size of substellar members is quite small, this would 
support the results from \cite{sch07} of mass-dependent disk lifetimes.

\cite{wya08} evaluates the protoplanetary disk fraction (based on near-IR excess) as a function of age for sun-like stars, and 
show that most sun-like stars have dispersed their disks by 6 Myr.  To extend 
this work into the M-type star regime, we cross-correlate with WISE the proposed members for 
the $\beta$ Pictoris association ($\sim$12 Myr) and $\eta$ Cha cluster ($\sim$5-8 Myr; \citealt{luh04}) from \cite{tor08} in the 
same manner as above for TWA, focusing solely on stars with spectral types between M0 and M6 (at this 
time, there are not enough substellar members of these similarly aged associations to 
probe any further than M6).  Using the same excess criteria as for TWA, we find that for the $\eta$ Cha cluster, of the 12 M-type stars listed in \cite{tor08}, 11 of which are individually detected with WISE, 5 show 22 $\mu$m excess.  For the $\beta$ Pic
association, we find only one marginal case of 22 $\mu$m excess out of 20 M-type members, coming from known debris disk-bearing
member AU Mic (\citealt{song02b}; \citealt{rhee07}).  In this spectral type range, we find a protoplanetary disk fraction of 
(4/19) $21^{+12}_{-6}$\% for TWA (5/20, or $25^{+11}_{-7}$\%, if possible members are included).  These results show a dramatic drop-off
of protoplanetary disk-fraction with age for M-stars, from $45^{+15}_{-3}$\% to $21^{+12}_{-6}$\% for  
$\eta$ Cha and TWA, to no evidence for protoplanetary disks in the $\beta$ Pic association.  This implies that protoplanetary disk 
dissipation in M-stars is a rapid process (occurring in a time-scale as short as a few Myr) with disk dispersal 
occurring sometime between 8 and 12 Myr.                          

\section{Conclusion}

Gathering the most recent relevant data, we have reassessed the current membership
status of the TW Hydrae association.  We find that there are 30 bona fide members.
Of these 30, 26 members were detected at 22 $\mu$m with WISE.  Using predicted 
stellar photosphere fluxes from SED fitting, we find 7 stellar and 3 substellar members
to show clear signs of IR-excess indicative of a circumstellar disk.  This is the first unambiguous evidence
of disks found for newly proposed members TWA 30A, TWA 31 and TWA 32, and the first mid-IR 
analysis of the disk around TWA 30B.  Our analysis shows a difference 
between the stellar and substellar disk fraction, evidence of mass-dependent disk lifetimes.  
We also note the dramatic drop-off of excess fraction in M0-M6 type stars when compared to the similarly aged
$\eta$ Cha cluster and the $\beta$ Pic association.

\acknowledgments

This research has made use of the SIMBAD database and VizieR catalog access tool, operated at CDS, 
Strasbourg, France.  This publication makes use of data products from the Two Micron All Sky Survey, 
which is a joint project of the University of Massachusetts and the Infrared Processing and 
Analysis Center/California Institute of Technology, funded by the National 
Aeronautics and Space Administration and the National Science Foundation, and the {\it Wide-field
Infrared Survey Explorer}, which is a joint project of the University of California, Los Angeles, and 
the Jet Propulsion Laboratory/California Institute of Technology, funded by the National 
Aeronautics and Space Administration.  We thank Adam J. Burgasser for a useful discussion.  C. M.
acknowledges support from the National Science Foundation under award No. AST-1003318.

\clearpage
\begin{deluxetable}{lccccccccccc}
\rotate
\tabletypesize{\scriptsize}
\tablecaption{TWA Membership}
\tablewidth{0pt}
\tablehead{
\colhead{TWA} & \colhead{SpT} & \colhead{dist} & \colhead{pmra} & \colhead{pmde} & \colhead{$V_{rad}$} & \colhead{Li} & \colhead{H$\alpha$} & \colhead{log($L_x$/$L_{bol}$)} & \colhead{$v$sin$i$} & \colhead{Member?} & \colhead{refs.}\\
\colhead{} & & \colhead{[pc]} & \colhead{mas/yr} & \colhead{mas/yr} & \colhead{km/s} & \colhead{m\AA} & \colhead{\AA} & \colhead{} &  km/s & &  }
\startdata
1 & K6Ve & 53 & -66.8 $\pm$ 1.6 & -15.2 $\pm$ 1.3 & 12.66 $\pm$ 0.22 & 435 & -172.5 & -2.59 & 6 & Y & 11,11,7,4,13,6,5,13\\ 
2A & M2Ve & 48 & -91.6 $\pm$ 1.8 & -20.1 $\pm$ 1.3 & 10.58 $\pm$ 0.51 & 520 & -1.72 & -3.54 & 13 & Y & 11,17,7,17,17,17,5,13\\
2B & M2 & 43 & \dots & \dots & \dots & \dots & \dots & \dots & \dots & N & 4,5,-,-,-,-,-,- \\ 
3A & M4Ve & 31 & -109.3 $\pm$ 8.7 & 0.8 $\pm$ 8.9 & 9.52 $\pm$ 0.86 & 510 & -40.89 & -3.33 & 12 & Y & 11,17,7,17,17,17,5,13\\
3B & M4Ve & 31 & \dots & \dots & 9.89 $\pm$ 0.62 & 540 & -4.26 & \dots & 12 & Y & 11,17,-,17,17,17,-,13\\
4A & K5V & 44 & -91.7 $\pm$ 1.5 & -30.0 $\pm$ 1.5 & 12.74 $\pm$ 0.10 & 380 & 0 & -3.43 & 9 & Y & 11,11,7,4,13,2,5,13\\
4B & K7,M1 & 47 & \dots & \dots & 5.73 $\pm$ 0.14 & 335,540 & 0 & -3.44 & \dots & Y & 4,5,-,4,4,2,5,-\\  
5A & M2Ve & 38 & -85.4 $\pm$ 3.6 & -23.3 $\pm$ 3.7 & 13.30 $\pm$ 2.00 & 630 & -6.37 & -3.05 & 54 & Y & 11,17,7,17,17,17,5,13\\
5B & M8Ve & 45 & -86 $\pm$ 8 & -21 $\pm$ 8 & 13.4 & 300 & -5.1 & -3.4 & 16 & Y & 11,11,9,1,13,1,16,13\\  
6 & K7 & 51 & -57.0 $\pm$ 2.1 & -20.9 $\pm$ 2.1 & 16.9 & 560 & -3.4 & -2.82 & 79.5 & Y? & 4,7,7,3,4,6,5,6\\  
7 & M2Ve & 34 & -122.3 $\pm$ 2.3 & -29.3 $\pm$ 2.3 & 12.21 $\pm$ 0.24 & 550 & -5.39 & -2.38 & 4 & Y & 11,17,7,17,17,17,5,13\\
8A & M3Ve & 44 & -99.3 $\pm$ 9.0 & -31.3 $\pm$ 8.9 & 8.34 $\pm$ 0.48 & 550 & -5.04 & -2.99 & 7 & Y & 11,17,7,17,17,17,2,13\\
8B & M5Ve & 27 & -95.3 $\pm$ 10.0 & -29.5 $\pm$ 10.3 & 8.93 $\pm$ 0.27 & 580 & -6.21 & \dots & 11 & Y & 11,17,7,17,17,17,-,13\\
9A & K5Ve & 68 & -52.8 $\pm$ 1.3 & -20.2 $\pm$ 1.8 & 9.46 $\pm$ 0.38 & 470 & -2.1 & -3.09 & 11 & Y & 11,11,7,4,13,6,5,13\\
9B & M1Ve & 68 & -70.7 $\pm$ 13.3 & -6.6 $\pm$ 15.8 & 11.3 & 480 & -4.3 & -2.28 & 8 & Y & 11,11,7,3,13,6,5,13\\  
10 & M2Ve & 67 & -72.6 $\pm$ 12.2 & -32.1 $\pm$ 12.3 & 6.75 $\pm$ 0.40 & 500 & -5.46 & -2.93 & 6 & Y & 11,17,7,17,17,17,5,13\\
11A & A0V & 67 & -53.3 $\pm$ 1.3 & -21.2 $\pm$ 1.1 & 9.4 $\pm$ 2.3 & 0 & -5.91 & \dots & 152 & Y & 11,11,7,4,13,5,-,13\\
11B & M2Ve & 67 & \dots & \dots & 9 $\pm$ 1 & 550 & -3.5 & -3.22 & 12 & Y & 11,11,-,4,13,6,5,13\\
11C & M4.5 & 67 & -49.6 $\pm$ 3 & -25.1 $\pm$ 3 &  \dots &  630 & -6.66 & \dots & \dots & Y & 19,19,19,-,19,19,-,-\\
12 & M2 & 63 & -36.3 $\pm$ 8.6 & -1.6 $\pm$ 8.9 & 13.12 $\pm$ 1.59 & 530 & -51.0 & \dots & 16.2 & Y? & 4,17,7,17,17,17,-,6\\ 
13A & M1Ve & 55 & -67.4 $\pm$ 11.8 & -17.0 $\pm$ 11.8 & 12.57 $\pm$ 0.50 & 580 & -3.0 & \dots & 12 & Y & 11,11,7,4,13,6,-,13\\
13B & M1Ve & 55 & \dots & \dots & 11.67 $\pm$ 0.64 & 550 & -3.0 & \dots & 12 & Y & 11,11, ,4,13,6,-,13\\
14 & M0 & 113 & -43.4 $\pm$ 2.6 & -7.0 $\pm$ 2.4 & 15.83 $\pm$ 2.00 & 590 & -5.68 & -3.15 & 43.1 & N? & 4,17,7,17,17,17,5,6\\
15A & M1.5 & 41 & -100.0 $\pm$ 33.0 & -16.0 $\pm$ 6.0 & 11.2 & 650 & -8.8 & -2.82 & 21.3 & N & 4,7,7,3,4,6,5,6\\  
15B & M2 & 100 & \dots & \dots & 10.03 $\pm$ 1.66 & 550 & -9.64 & \dots & 32.3 & ? & 4,17,-,17,17,17,-,7\\
16 & M1Ve & 65 & -53.3 $\pm$ 5.2 & -19.0 $\pm$ 5.2 & 9.01 $\pm$ 0.42 & 380 & -3.08 & -3.43 & 11 & Y & 11,17,7,17,17,17,5,13\\
17 & K5 & 163 & -28.0 $\pm$ 8.5 & -11.1 $\pm$ 8.5 & 4.6 & 490 & -3.2 & -3.23 & 49.7 & N & 4,7,7,3,4,6,5,6\\  
18 & M0.5 & 121 & -29.0 $\pm$ 5.2 & -21.2 $\pm$ 5.2 & 6.9 & 420 & -3.3 & -3.29 & 24.1 & N & 4,7,7,3,4,6,5,6\\  
19A & G5 & 109 & -33.6 $\pm$ 0.9 & 8.5 $\pm$ 0.9 & 11.5 $\pm$ 3.8 & 190 & 0.57 & -3.48 & 25 & N & 4,7,7,4,4,2,5,5\\
19B & K7 & 103 & -35.6 $\pm$ 4.8 & -7.5 $\pm$ 4.6 & 15.2 & 400 & -2.2 & -2.99 & 48.7 & N & 4,7,7,2,4,6,5,6\\  
20 & M3Ve & 73 & -52.0 $\pm$ 5.0 & -16.0 $\pm$ 6.0 & 8.1 $\pm$ 4 & 160 & -3.1 & -3.09 & 30 & Y & 11,11,7,2,13,6,5,13\\
21 & M1 & 45 & -65.3 $\pm$ 2.4 & 13.7 $\pm$ 1.0 & 17.5 $\pm$ 0.8 & 290 & 0.0 & -3.59 & \dots & N? & 5,7,7,3,3,3,5,-\\   
22 & M5 & 22 & -175.8 $\pm$ 0.8 & -21.3 $\pm$ 0.8 & 13.57 $\pm$ 0.26 & 650 & -10.48 & \dots & 9.7 & Y? & 6,17,12,17,17,17,-,17\\ 
23 & M1 & 61 & -68.0 $\pm$ 4.0 & -23.0 $\pm$ 4.0 & 8.52 $\pm$ 1.20 & 500 & 8.12 & -2.94 & 14.8 & Y & 6,17,7,17,17,17,5,6\\
24 & K3 & 107 & -34.4 $\pm$ 2.8 & -13.1 $\pm$ 1.7 & 11.9 $\pm$ 0.9 & 340 & -0.3 & -3.16 & 13.0 & N & 6,7,7,3,3,6,5,6\\
25 & M1Ve & 51 & -75.0 $\pm$ 2.0 & -26.9 $\pm$ 1.4 & 9.2 $\pm$ 2.1 & 555 & -2.4 & -2.95 & 13 & Y & 11,11,7,3,13,6,5,13\\
26 & M8Ve & 41 & -93 $\pm$ 5 & -31 $\pm$ 10 & 11.6 $\pm$ 2 & 500 & -44.7 & -4.8 & 25 & Y & 11,11,9,1,13,8,16,13\\
27 & M8Ve & 53 & -63 $\pm$ 3 & -23 $\pm$ 2 & 11.2 $\pm$ 2 & 500 & -10.2 & $<$-4.8 & 13 & Y & 11,11,9,1,13,8,16,13\\
28 & M8.5 & 55.2 & -67.2 $\pm$ 0.6 & -14.0 $\pm$ 0.6 & \dots & \dots & -64 $\pm$ 3 & $<$-4.0 & \dots & Y & 10,10,10,-,-,9,16,-\\ 
29 & M9.5 & 90 & -89.4 $\pm$ 10 & -20.9 $\pm$ 10 & \dots & \dots & -15 $\pm$ 3 & \dots & \dots & Y & 9,20,9,-,-,9,-,-\\   
30A & M5 & 42 & -89.6 $\pm$ 1.3 & -25.8 $\pm$ 1.3 & 12.3 $\pm$ 1.5 & 610 & -6.8 & -3.34 & \dots & Y & 14,14,14,14,14,14,14,-\\   
30B & M4 & 44 & -83 $\pm$ 9 & -30 $\pm$ 9 & 12 $\pm$ 3 & 500 & -7.4 & \dots & \dots & Y & 15,15,15,15,15,15,-,- \\ 
31 & M4.2 & 110 & -42 $\pm$ 6 & -36 $\pm$ 3 & 10.47 $\pm$ 0.41 & 410 & -114.8 & \dots & \dots & Y? & 17,17,17,17,17,17,-,-\\
32 & M6.3 & 53 & -62.2 $\pm$ 3.5 & -24.7 $\pm$ 3.9 & 7.15 $\pm$ 0.26 & 600 & -12.6 & $<$-3.10 & \dots & Y & 17,17,17,17,17,17,18,- \\ 
\enddata
\\
The references in column 12 refer to columns 2, 3, 4 and 5, 6, 7, 8, 9, and 10, respectively.\\  
References: (1) \cite{moh03}; (2)  \cite{reid03};
 (3) \cite{song03}; (4) \cite{tor03}; (5) \cite{reza04}; (6) \cite{jay06};
 (7) \cite{mam05}; (8) \cite{bar06}; (9) \cite{loop07}; (10) \cite{teix08}; (11) \cite{tor08} 
 (12) \cite{teix09}; (13) \cite{das09}; (14) \cite{loop10a}; (15) \cite{loop10b};
 (16) \cite{cas11}; (17) \cite{shk11}; (18) \cite{rod11}; (19) \cite{kas08}; (20) This work.
 \end{deluxetable}
\begin{deluxetable}{lcccccccccccc}
\rotate
\tabletypesize{\scriptsize}
\tablecaption{TWA Excess Measurements\tablenotemark{a}}
\tablewidth{0pt}
\tablehead{
\colhead{TWA} & & \colhead{W1} & \colhead{mJy} & & & \colhead{W3} & \colhead{mJy} & & & \colhead{W4} & \colhead{mJy}\\
\colhead{} & \colhead{Phot} & \colhead{Meas} & \colhead{Err} & \colhead{Ex} & \colhead{Phot} & \colhead{Meas} & \colhead{Err} & \colhead{Ex} & \colhead{Phot} & \colhead{Meas} & \colhead{Err} & \colhead{Ex} }
\startdata
 1   & 366.7 & 447.0 & 13.4 & 6.0 & 34.5 & 486.5 & 6.2 & 72.5 & 11.0 & 2070.0 & 30.3 & 68.0\\
 2   & 710.9 & 685.4 & 23.6 & -1.1 & 77.1 & 81.9 & 1.2 & 4.0 & 25.8 & 24.5 & 1.1 & -1.2\\
 3AB   & 778.5 & 708.5 & 24.4 & -2.9 & 98.1 & 891.9 & 13.0 & 60.8 & 33.3 & 1693.4 & 21.7 & 76.5\\
 4AB   & 1912.2 & 1976.6 & 109.7 & 0.6 & 179.8 & 1807.6 & 19.9 & 81.9 & 57.4 & 6994.6 & 64.1 & 108.2\\
 5AB   & 710.9 & 674.7 & 23.2 & -1.6 & 77.1 & 86.7 & 1.3 & 7.6 & 25.8 & 24.9 & 1.1 & -0.8\\
 6   & 216.2 & 224.2 & 4.9 & 1.6 & 21.5 & 25.9 & 0.4 & 9.9 & 7.1 & 6.4 & 0.8 & -0.8\\
 7   & 576.6 & 581.2 & 15.8 & 0.3 & 57.3 & 72.9 & 1.0 & 15.6 & 18.9 & 33.1 & 1.2 & 12.2\\
 8A  & 359.4 & 350.8 & 9.2 & -0.9 & 40.0 & 44.5 & 0.5 & 9.0 & 13.4 & 12.8 & 1.0 & -0.6\\
 8B  & 101.0 & 88.3 & 4.8 & -2.6 & 12.4 & 13.4 & 0.5 & 2.0 & 4.2 & 2.9 & 0.8 & -1.4\\
 9A  & 259.7 & 265.7 & 6.8 & 0.8 & 24.8 & 29.7 & 0.5 & 9.8 & 8.2 & 9.1 & 1.0 & 0.9\\
 9B  & 83.0 & 77.2 & 3.5 & -1.7 & 9.8 & 9.5 & 0.4 & -0.8 & 3.29 & $<$3.8 & \dots & \dots \\
 10  & 184.9 & 180.4 & 3.8 & -1.2 & 21.8 & 22.6 & 0.4 & 2.0 & 7.3 & 6.7 & 0.6 & -1.0\\
 11AB & 1442.3 & 2028.2 & 112.6 & 5.2 & 123.9 & 278.4 & 3.8 & 40.4 & 36.2 & 2713.7 & 34.8 & 77.0\\
 11C & 107.6 & 93.8 & 1.9 & -7.2 & 15.5 & 14.5 & 0.3 & -3.3 & 5.4 & 6.3 & 0.6 & 1.5\\
 12  & 210.9 & 187.2 & 3.9 & -6.1 & 21.6 & 23.7 & 0.4 & 5.25 & 7.2 & 5.2 & 0.8 & -2.5\\
 13A & 294.3 & 273.3 & 12.8 & -1.64 & 31.1 & 33.9 & 0.8 & 3.5 & 10.3 & 13.3 & 1.6 & 1.9\\
 13B & 350.2 & 326.5 & 16.1 & -1.47 & 37.0 & 35.4 & 0.8 & -2.0 & 12.3 & 8.2 & 1.5 & -2.7\\
 16  & 212.7 & 206.0 & 4.3 & -1.6 & 23.7 & 25.4 & 0.4 & 4.3 & 7.9 & 8.0 & 0.7 & 0.1\\
 20  & 149.8 & 145.0 & 3.0 & -1.6 & 17.7 & 18.5 & 0.3 & 2.7 & 6.0 & 6.6 & 0.6 & 1.0\\
 22 & 327.1 & 311.0 & 6.5 & -2.5 & 41.2 & 46.0 & 0.6 & 8.0 & 14.0 & 15.8 & 0.8 & 2.3\\
 23  & 288.9 & 271.6 & 6.4 & -2.7 & 34.1 & 34.5 & 0.5 & 0.8 & 11.5 & 10.7 & 0.8 & -1.0\\
 25  & 392.3 & 384.7 & 10.1 & -0.8 & 41.4 & 44.6 & 0.7 & 4.5 & 13.8 & 12.3 & 0.8 & -1.9\\
 26\tablenotemark{b}  & 11.1 & 10.7 & 0.2 & -2.0 & 1.7 & 1.8 & 0.1 & 0.9 & 0.58 & 0.59 & 0.01 & 0.1\\
 27  & 7.1 & 7.4 & 0.2 & 1.5 & 0.9 & 5.2 & 0.1 & 34.0 & 0.3 & 5.1 & 0.6 & 8.2\\
 28  & 9.0 & 8.3 & 0.2 & -3.5 & 1.4 & 5.6 & 0.2 & 21.0 & 0.5 & 5.2 & 0.8 & 5.9\\
 29  & 1.6 & 2.0 & 0.1 & 4.0 & 0.2 & 0.3 & 0.3 & 0.3 & 0.1 & $<$1.9 & \dots & \dots \\
 30A & 115.3 & 93.8 & 1.9 & -11.3 & 14.1 & 47.0 & 0.7 & 47.0 & 4.8 & 72.4 & 1.8 & 37.6\\
 30B\tablenotemark{c} & 132.4 & 3.7 & 0.1 & -1287.0 & 18.1 & 36.5 & 0.6 & 30.7 & 6.29 & 78.4 & 1.9 & 37.9\\
 31  & 6.0 & 6.2 & 0.1 & 1.0 & 0.8 & 3.3 & 0.1 & 20.3 & 0.3 & 6.1 & 0.7 & 8.2\\
 32  & 53.5 & 46.6 & 1.0 & -6.8 & 7.7 & 24.3 & 0.4 & 43.9 & 2.7 & 40.5 & 1.4 & 27.9\\
\enddata
\tablenotetext{a}{For late type stars, the spectral models in the WISE channel 2 region are affected by the fundamental CO bandhead at 4.7 $\mu$m, so are not included here.  No color correction has been applied to the WISE data.}
\tablenotetext{b}{The measured and estimated fluxes for TWA 26 in the W4 columns correspond to its MIPS 24 $\mu$m measurement.} 
\tablenotetext{c}{The estimated flux values for TWA 30 are from the distance-scaled spectral model (see Figure 1 and the discussion Section 3.} 
\end{deluxetable}
\begin{figure}
\plotone{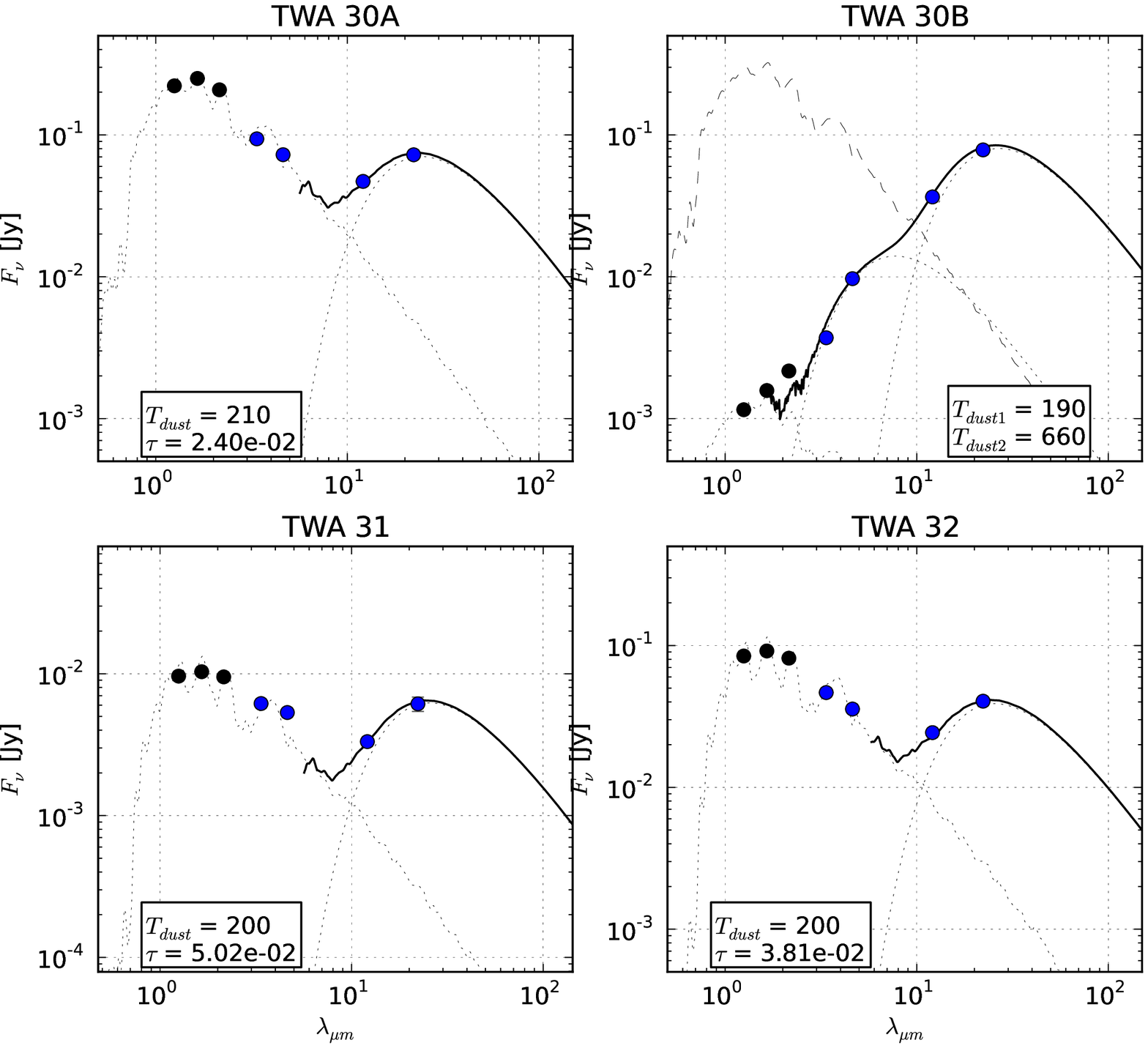}
\caption{Spectral energy distributions (SED) for the four newest proposed members of TWA.  The dotted curves are the spectral model and the single temperature blackbody dust fit.  The solid curve is the spectral model $+$ blackbody dust fit.  For TWA 30B, the dashed line indicates the M4 model spectrum scaled to the distance quoted in Table 1 (see discussion in Section 3).  The other curves represent the best fit to the currently available data as for TWA 30A, 31, and 32.  To match the observed TWA 30B data points, we attempted to fit the SED with a combination of a reddened stellar spectrum using the standard ISM extinction relation \citep{mat90} and gray extinction, but could not find a suitable fit.  This implies that the dust grain size distribution around TWA 30B is different from that of ISM grains.}
\end{figure}
\begin{figure}
\plotone{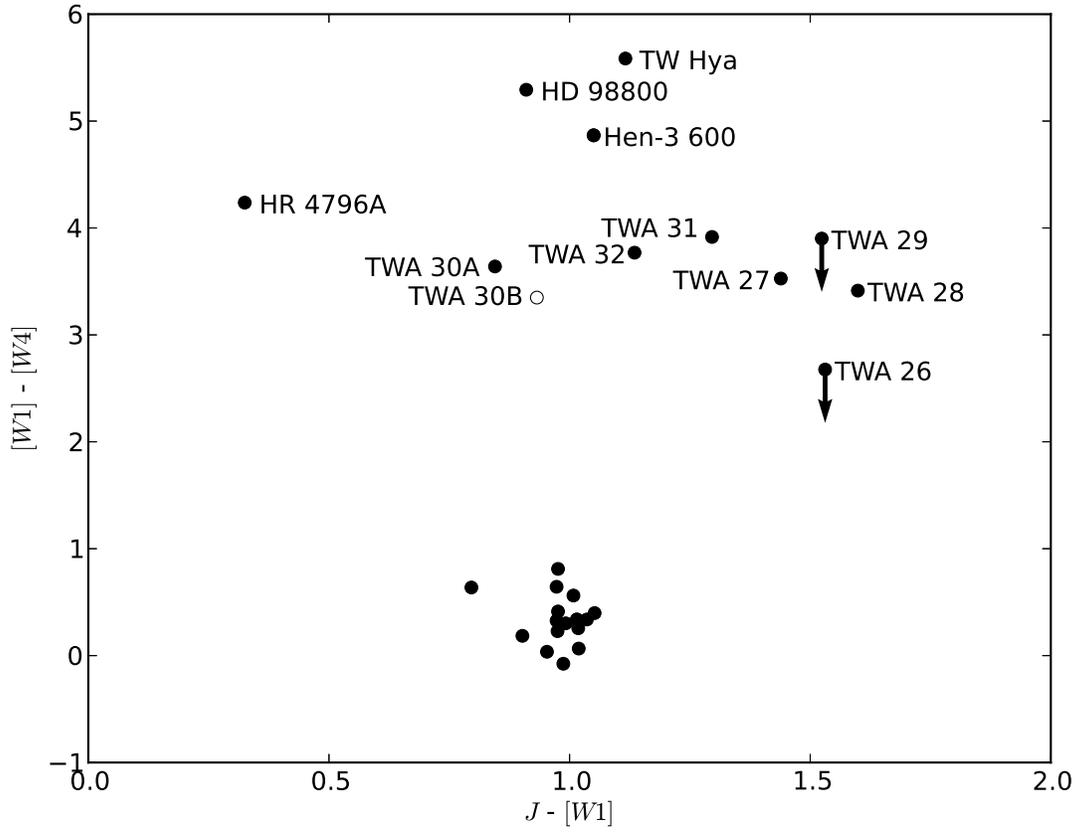}
\caption{J-W1 vs. W1-W4 color-color diagram for all individually detected TWA true and possible members.  A clear distinction can be seen between those sources harboring disks and those without.  For TWA 30B (open symbol), the J and W1 magnitudes are determined from the spectral model, while W4 is the measured magnitude.  The actual position of TWA 30B on this figure from its measured data is [2.2,7.2].}
\end{figure}

\clearpage

\end{document}